\documentclass{aa}
\usepackage{txfonts}
\usepackage{graphicx}

\begin{document}

\title{On the tidal tails of Milky Way globular clusters}

\author{Andr\'es E. Piatti\inst{1,2}\thanks{\email{andres.piatti@unc.edu.ar}} and Julio A. Carballo-Bello\inst{3} 
}

\institute{Instituto Interdisciplinario de Ciencias B\'asicas (ICB), CONICET UNCUYO, Padre J. Contreras 1300, (5500) Mendoza, Argentina;
\and Consejo Nacional de Investigaciones Cient\'{\i}ficas y T\'ecnicas (CONICET), Godoy Cruz 2290, C1425FQB,  Buenos Aires, Argentina;
\and Instituto de Astrof\'{\i}sica, Facultad de F\'{\i}sica, Pontificia Universidad Cat\'olica 
de Chile, Av. Vicu\~na Mackenna, 4860, 782-0436, \\ Macul, Santiago, Chile\\
}

\date{Received / Accepted}

\abstract{
We report on the search for overall kinematical or structural conditions that have
allowed some Milky Way globular clusters to presently develop tidal tails. For this purpose, 
we build a comprehensive catalogue of globular clusters with studies focused on
their outermost regions and classified them in three categories: those with
observed tidal tails, those with extra-tidal features different from tidal tails
and those without any signature of extended stellar density profiles.
When exploring different kinematical and structural parameter spaces, we
found that globular clusters - irrespective from the presence of tidal
tails, or any other kind of extra-tidal features or the absence of them - behave 
similarly. In general,  globular clusters whose orbits are relatively more eccentric
and very inclined respect to the Milky Way plane have undergone a larger amount of
mass-loss by tidal disruption. The latter has also accelerated the
internal dynamics toward a comparatively more advanced stage of evolution. 
These outcomes show that it is not straightforward to find any particular set of parameter space and dynamical conditions that can definitely predict tidal tails along globular clusters in the Milky Way.
}
 
 \keywords{
Galaxy: globular clusters: general -- 
Methods: observational.}

\titlerunning{Tidal tails of globular clusters}

\authorrunning{A.E. Piatti and J.A. Carballo-Bello}

\maketitle

\markboth{A.E. Piatti and J.A. Carballo-Bello: }{Tidal tails of globular clusters}

\section{Introduction}

The formation of stellar streams or tidal tails due to the disruption or dissolution of Galactic
globular clusters has long been understood as a consequence of their interaction with 
their host galaxy \citep{grillmairetal1995,dehnenetal2004}. Indeed, \citet{montuorietal2007} performed detailed N-body simulations
to show that tidal  tails are generated in globular clusters as a consequence of their strong 
interaction with the densest components of the Galaxy, e.g. the bulge and the disk, which 
may result in the presence of multiple tidal tails after the repetitive apocenter passages
 \citep{hb2015}. However, rather than from tidal shocks,
\citet{kupperetal2010,kupperetal2012} analytically and numerically showed that tidal
tails and their substructures are originated from the epicyclic motions of a continuous 
stream of stars escaping the clusters, regardless whether the clusters' orbits are circular 
or eccentric.

From an observational point of view, there has been a number of studies on the outermost 
regions of globular clusters with the purpose of finding out tidal tails 
\citep[see, e.g.][]{malhanetal2018,mateuetal2018,pme2019}. The outcomes have 
been dissimilar, since some clusters have been found to have tidal tails, others azimuthally 
(position angle with respect to the globular cluster centers) irregular extended halos or clumpy structures \citep[see, e.g.][]{moore1996,ibataetal2013,kuzmaetal2016}, or simply \citet{king62}'s radial profiles 
without extra-tidal features (see Table~\ref{tab:tab1}). Precisely, in this work we carry out a 
comprehensive  compilation of these relevant observational results obtained up-to-date to
investigate whether there is any cluster  structural or internal  dynamical properties,
and/or orbital parameters that allow us to identrify globular clusters with tidal tails. 

This work is organized as follows: in Section 2 we build a catalogue of Milky Way
globular clusters with reliable studies on their outermost structures. From this
catalogue we analyze in Section 3 different parameter spaces, including kinematical, 
positional, structural, internal dynamics properties, in order to disentangle any
general conditions related to the presence of tidal tails in Milky Way globular clusters.
Finally, in Section 4 we summarize the main conclusions of this work.

\section{The catalogue of extra-tidal features}

We searched the literature looking for studies focused on the structure of the 
outermost regions of Milky Way globular clusters. In this work, we classify the clusters as 
G1, G2, and G3, this is, globular clusters with tidal tails, clusters with extra-tidal features
(those that are not symmetric tails, but distributed around the globulars
clusters), 
and those without any signature of extended structures, respectively.
 In G1, we included those clusters with clear tails 
extended  beyond the cluster's Jacobi radius. Since the Jacobi radius changes between  the 
perigalactic and apogalactic positions, we used the values computed by \citet{piattietal2019b} 
for the sem-major axis of the cluster's orbit (their equations 1 and 3). Clusters with extra-tidal 
structures (e.g., azimuthally irregular clumpy features with respect to the globular cluster center, extended halos) are included in the 
G2 group,  while those clusters with stellar radial profiles that do not show any 
excess beyond the tidal radius given by the \citet{king62}'s model are gathered into the 
G3 group. We would like to note that distinction between G1 and G2 groups relies on the shape of their
extra-tidal features: while for G1 clusters extra-tidal stars are distributed along symmetric tails,
those of G2 clusters are spread around the clusters, in halos more or less uniformly
populated. Such a difference is readily visible while inspecting the stellar density maps and
density profiles as a function of the position angle measured from the cluster centers
(see references in Table~\ref{tab:tab1}).

\begin{table*}
\caption{Relevant references from the literature of clusters in the G1, G2 and G3 groups.}
\label{tab:tab1}
\begin{tabular}{@{}lccclccclccc}\hline\hline

\hspace{0.cm} ID   & G1 Ref.  &  G2 Ref.  & G3 Ref.  &  \hspace{0.cm} ID  & G1 Ref.  &  G2 Ref.  & G3 Ref.  & \hspace{0.cm} ID  & G1 Ref.  &  G2 Ref.  & G3 Ref. \\
\hline 
NGC\,104    &              &   6    &        &    NGC\,5694 &      &  17    &         & NGC\,7089     &       &            &     19   \\
NGC\,288    &    1,10   &       &        &    NGC\,5824   &    &          &   30   & NGC\,7492     &  22  &         &        \\
NGC\,362    &   25       &         &        &   NGC\,5904   &  29 &      &        &   AM\,4             &       &            &      14  \\
NGC\,1261  &              &  30   &        &    NGC\,6205  &     &          &   19   &  Crater            &       &            &      11   \\ 
NGC\,1851  &   10,20  &         &        &    NGC\,6229  &      &         &   14   &  Eridanus       & 23    &           &         \\     
NGC\,1904  &             &   20    &       &    NGC\,6266  &     &   13   &        &  ESO\,452-SC11 &     &           &       7   \\
NGC\,2298  &             &   18,20    &       &    NGC\,6273  &     &   13    &        &  FSR\,1758        &       &          &      2   \\   
NGC\,2419  &             &  19     &       &    NGC\,6341  &     &          &   19   & Liller\,1             &         &          &     12  \\  
NGC\,2808  &             &  20     &       &    NGC\,6362  &     &   27    &        &  Pal\,1              &  24    &          &        \\     
NGC\,3201  &             &  15     &       &    NGC\,6544  &     &   16    &        &  Pal\,3               &        &         &       19  \\  
NGC\,4147  &    19     &        &        &     NGC\,6626 &     &   13    &        &  Pal\,4               &        &         &       19  \\    
NGC\,4590 &     26     &          &       &     NGC\,6642 &     &   13   &        & Pal\,5                & 5,28  &        &         \\  
NGC\,5024  &             &          &   14,19  &     NGC\,6656 &     &   15   &        &  Pal\,12              &        &        &          3 \\     
NGC\,5053  &             & 19     &        &     NGC\,6681 &     &     8    &        &  Pal\,14              &   32   &        &        \\       
NGC\,5139  &     4,21 &       &       &     NGC\,6779 &      &     9   &        &  Pal\,15              & 23    &        &        \\     
NGC\,5272  &             &          &   14,19  &     NGC\,6864 &     &          &   14   &  Rup\,106           &        &        &        14 \\   
NGC\,5466  &     19,31   &        &        &     NGC\,7006 &      &    19  &        & Whiting\,1          &         &    14    &        \\       
NGC\,5634  &              &  14   &        &    NGC\,7078  &      &    19  &        &                        &.          &.           &             \\
\hline
\end{tabular}

\noindent Ref: (1) \citet{kaderalietal2019}; (2) \citet{barbaetal2019}; (3) \citet{musellaetal2018}; (4) \citet{s2019}; 
(5) \citet{odenetal2001}; (6) \citet{p17c};  (7) \citet{kochetal2017};  (8) \citet{hanetal2017}; (9) \citet{pc2019};
(10) \citet{shippetal2018}; (11) \citet{weiszetal2016};  (12) \citet{saracinoetal2015}; (13) \citet{chunetal2015}; 
(14) \citet{carballobelloetal2014}; (15) \citet{kunderetal2014}; (16) \citet{cohenetal2014}; (17) \citet{correntietal2011}; 
(18) \citet{balbinotetal2011}; (19) \citet{jg2010}; (20) \citet{carballobelloetal2018}; (21) \citet{ibataetal2019}; 
(22) \citet{naverreteetal2017}; (23) \citet{myeongetal2017}; (24) \citet{noetal2010}; (25) \citet{carballobello2019}; 
(26) \citet{pme2019}; (27) \citet{kunduetal2019}; (28) \citet{starkmanetal2019}; (29) \citet{g2019}; (30) \citet{kuzmaetal2018};
(31) \citet{belokurovetal2006}; (32) \citet{sollimaetal2011}.
\end{table*}

We found 53 globular clusters with reliable structural information, which represent $\approx$
1/3 of those included in the \citet[][2010 Edition]{harris1996}'s catalogue. Although
the cluster sample of Table~\ref{tab:tab1} is not complete, it results useful to find any
intrinsic difference between the properties of clusters in the three defined categories. These, 
in turn, can shed light on our knowledge about the different modes of cluster dissolution.
Table~\ref{tab:tab1} does not list every published paper on this field, so that
pioneer works surpassed by recent analysis have been omitted. 

Table~\ref{tab:tab1} contains 14, 22 and 17 clusters in groups G1, G2 and G3, respectively.
Table~\ref{tab:tab1} would seem to suggest that it is really hard to detect debris tails around globular clusters. Nevertheless, we do not know if every globular cluster should have tidal 
tails. Globular clusters orbiting very far from the Galactic center will not experience huge tidal forces of the Galactic potential, and will probably never develop stellar debris. Likewise,
this observational
evidence poses the question about what conditions may favor a globular cluster to have
extra-tidal features. For instance, we can ask whether different kinematical histories (orbits) make a
difference to this respect, or whether cluster properties (e.g., size,  mass)  within certain 
values are correlated with the existence of tidal tails, among others. 

\section{Analysis and discussion}

We started exploring whether the  orbital history of globular clusters is related
to the occurrence of tidal tails. To this respect, we followed the analysis of 
\citet{piatti2019}, who found that within the most frequently used  orbital properties, 
the space defined by the eccentricity, the inclination of the orbit and the  semi-major axis
($a$) turns out to be the best enlightenment of the overall   orbital state of the globular 
cluster system. We added to our analysis the ratio of the  cluster mass lost by disruption
to the  total initial cluster mass  ($M_{dis}/M_{ini}$) computed by \citet{piattietal2019b},
to study at what 
extent the Milky Way gravitational field has shaped the structural parameters and 
internal dynamics of its globular cluster population.    \citet{piattietal2019b} estimated how much clusters have been disrupted due to relaxation and tidal heating, 
and split the difference between the initial mass $M_{ini}$ and the current mass $M_{GC}$ 
- both taken from  \citet{baumgardtetal2019} - up between mass lost via stellar evolution ($M_{ev}$) and mass lost due to disruption ($M_{dis}$):
 
 \begin{equation}
 M_{ini} = M_{GC} + M_{ev} + M_{dis}, 
 \end{equation}
  
 \noindent with $M_{ev}$ = 0.5$\times$$M_{ini}$, from which they got:
 
\begin{equation}
M_{dis}/M_{ini} = 1/2  - M_{GC}/M_{ini} .
\end{equation}

\noindent  $M_{ini}$ values in \citet{baumgardtetal2019} were obtained by integrating 
each cluster's orbit backwards in time from their observed positions and space velocities 
and measured current  masses, taking into consideration the dynamical drag force. 
 It was additionally assumed that clusters lose half of their
$M_{ini}$ due to stellar evolution during their first Gigayear. They iterated over a wide range of $M_{ini}$ values  until they were able to recover each cluster's $M_{GC}$, on the basis of a linear mass loss dependence with time in a spherically symmetric, isothermal galaxy potential over the entire age of each cluster.

\begin{figure}
\includegraphics[width=\columnwidth]{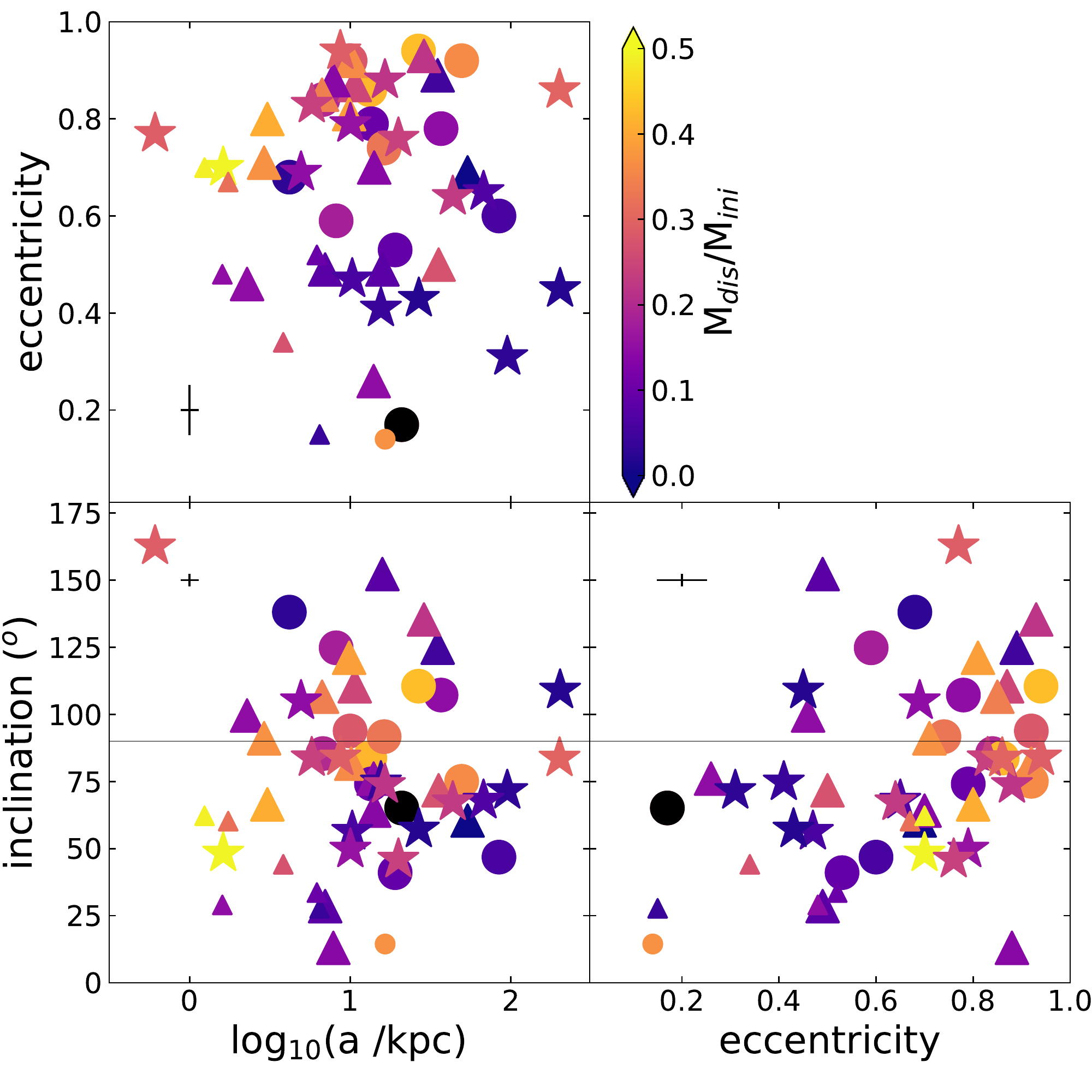}
\caption{Relationship between the semimajor axis, the eccentricity and the orbital
inclination for globular clusters in groups G1, G2 and G3, drawn with filled circles, 
triangles and stars, respectively. Large and small symbols refer to clusters with an 
accreted origin or formed {\it in-situ}, respectively, according to \citet{massarietal2019}.
The horizontal line in the bottom panels (inclination = 90$\degr$) splits them into the 
prograde (inclination < 90$\degr$) and retrograde (inclination > 90$\degr$) regimes 
(see \citet{piatti2019}). Colour bar  represents the ratio of the cluster mass lost by
disruption to the total initial mass (see \citet{piattietal2019b}). Error bars are also
included. Pal\,5 is represented by a large black filled circle.}
\label{fig:fig1}
\end{figure}

Fig.~\ref{fig:fig1} shows the relationship between the aforementioned globular
cluster parameters. We differentiated clusters in groups G1, G2 and G3 by
representing them with filled circles, triangles and stars, respectively. The ratio
of the cluster mass lost by disruption to the total cluster mass was used to color
the filled symbols as indicated by the adjacent bar. 
The convention for the orbital inclination is as follows: clusters rotating in prograde
orbits, i.e, in the direction of the Milky Way rotation, have orbital inclinations
$<$ 90$\degr$; those in retrograde orbits have inclinations $>$ 90$\degr$.
We additionally considered the origin of the globular clusters according to
\citet{massarietal2019}, namely: clusters associated with an accreted dwarf
galaxy (larger symbols) or formed {\it in-situ} (smaller symbols). We found 1, 6 and
0 clusters formed {\it in-situ} in groups G1, G2 and G3, respectively.

It is easily derived from Fig.~\ref{fig:fig1} that, in general terms, there is no clues 
for distinguishing globular clusters having tidal tails. Indeed, it would seem that all three 
defined groups contain  clusters spread over approximately similar ranges of 
eccentricity, inclination and semi-major axis. From this point of view, tidal tails
would not seem to arise from the overall kinematical pattern of the globular clusters.
However, and as can be seen, any globular cluster moving in an orbit with a relatively large
eccentricity, i.e., along a
more radial orbit, has lost a higher amount of its initial mass due to
tidal disruption, as compared to those with smaller eccentricities. Such a behaviour is 
observed in bulge (log($a$ /kpc) $\le$ 0.5), disk  (0.5 $<$ log($a$ /kpc)  $\le$ 1.3)) and 
outer halo  (log($a$ /kpc) $>$ 1.3) globular clusters. Nevertheless, a  high
eccentricity alone would not seem to be enough to produce a large amount of 
mass-loss (see top left panel of Fig.~\ref{fig:fig1}).

When a relatively high eccentricity ($\ga$ 0.8) is combined with a very inclined orbit
($|$inclination - 90$\degr$$|$ $\la$ 20$\degr$), it is possible to isolate a group of
clusters - irrespective from groups G1, G2 and G3 - with a relatively large amount of 
mass lost by disruption (see bottom-right panel of Fig.~\ref{fig:fig1}). Seemingly, 
relatively large eccentricities 
and low orbital inclinations or relatively very inclined orbits and low eccentricities
are less efficient in terms of cluster mass tidal disruption. 
\citet[][and reference therein]{webbetal2014} showed that repeated disk passages
can contribute to the cluster mass loss by disruption. Hence, a possible scenario that
would increase the chances of repeated disk crossing is that of clusters with 
relatively high eccentricity/inclination values \citep[see][]{piatti2019}, which could
explain the loci of clusters with disrupted mass larger than $\sim$ 0.3 in the
bottom-right panel of Fig.~\ref{fig:fig1}.

\begin{figure}
\includegraphics[width=\columnwidth]{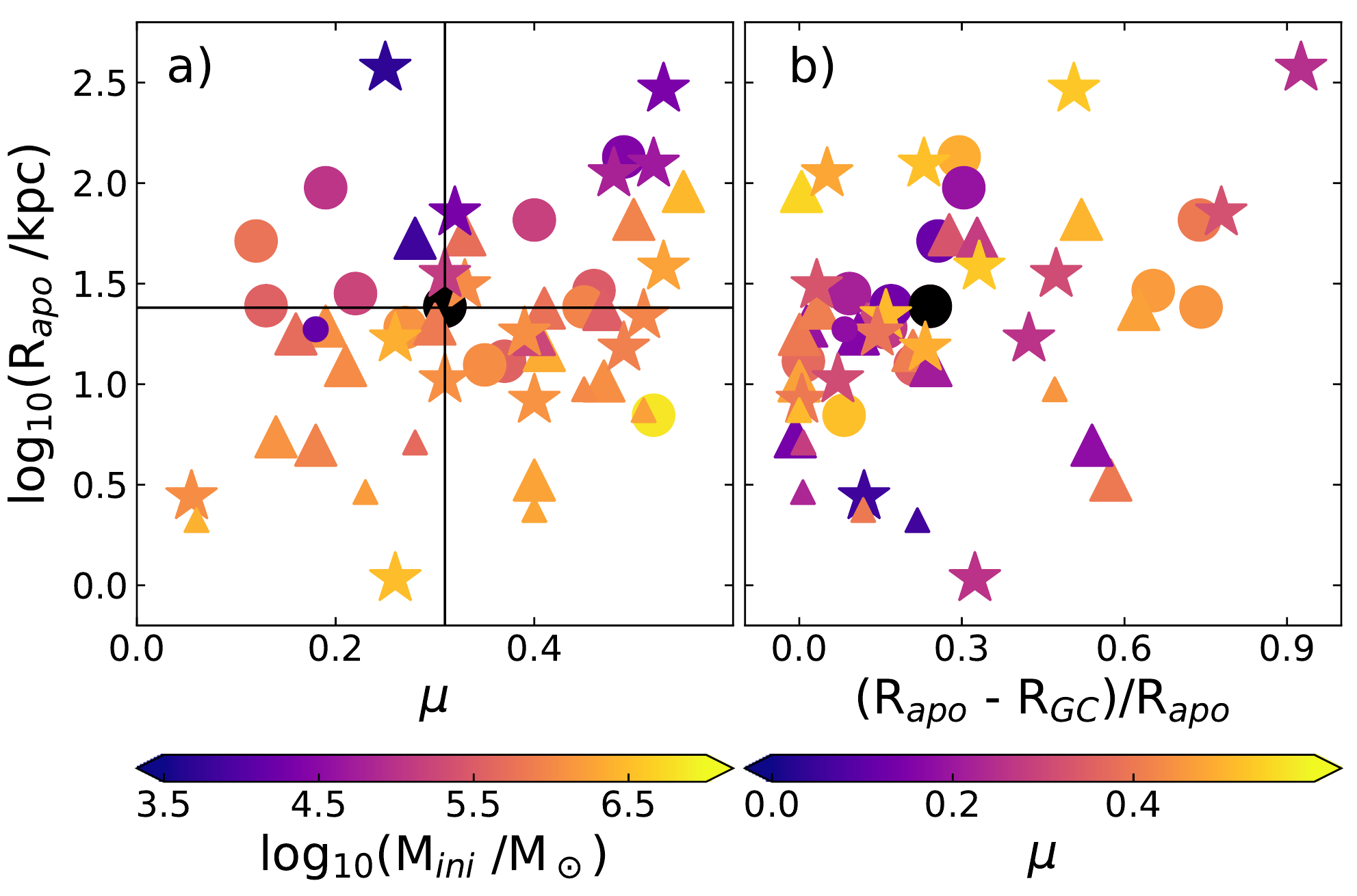}
\caption{Relationships between different position and mass-related parameters
for the globular clusters in groups G1, G2 and G3. Symbols are as in Fig.~\ref{fig:fig1}.
Panel a is divided in four quadrants as in figure 5 of \citet{bg2018}.
 Pal\,5 is represented by a large black filled circle.}
\label{fig:fig2}
\end{figure}

\citet{bg2018}  explored the formation of tidal tails around Milky Way globular clusters 
from the combination of the fast cluster evolution code Evolve Me A Cluster of StarS 
\citep[EMACS, ][]{ag2012}, a semi-analytical model for the evolution of the stellar 
mass function and a fast stream simulation code. They found that globular clusters 
with tidal tails are close to dissolution and also likely close to their apogalacticon.
We reproduced their figure 5 (see Fig.~\ref{fig:fig2},a) using the apogalactic distances
($R_{apo}$) from \citet{baumgardtetal2019} and the remaining mass 
fraction  ($\mu$= 1 - $M_{dis}/M_{ini}$)  from equations 4 and 5 of
 \citet{piattietal2019b}, which rely on the actual and initial cluster masses computed
 by  \citet{baumgardtetal2019}. For the sake of
the reader, we used the same symbols as in Fig.~\ref{fig:fig1} and subdivided
the figure in four panels as in \citet{bg2018}.

According to \citet{bg2018}, the upper-left panel of Fig.~\ref{fig:fig2},a encompasses 
$R_{apo}$ and $\mu$ values more extreme that those for Pal\,5, so that globular 
clusters placed there should be good candidates to develop tidal tails. As can be 
seen, some globular clusters with observed tidal tails are distributed in that panel, 
alongside clusters that exhibit extra-tidal structures (G2 group clusters) and also 
AM\,4 at $\approx$ (0.25,2.6), which belong to the G3 group. Therefore, 
extreme $R_{apo}$ and $\mu$ values would not be exclusive of G1 group
globular clusters. On the other hand, globular clusters with tidal tails also occupy the 
right-hand panels, i.e., those with comparatively smaller amounts of mass loss 
by tidal disruption. This result suggests that tidal tails can develop early in the
cluster disruption process.  As for the
closeness of their present galactocentric positions ($R_{GC}$) to the respective apogalacticon, 
Fig~\ref{fig:fig2},b would seem to show that such a condition is not 
verified by the observations. Indeed, G1 group clusters can have a remaining mass
fraction larger than 0.4 and be located reasonably far from their apogalacticon
(($R_{apo}-R_{GC}$)/$R_{apo}$ > 0.6).
Additionally we note that globular clusters  with tidal tails are not necessarily those 
initially more massive (see Fig.~\ref{fig:fig2},a), although it is known from theory and 
numerical simulations that the mass-loss rate is a function of the form $\dot{M}$ 
$\propto$ $-M^{1/4}/R_{GC}$ for globular clusters evolving in an isothermal halo
\citep{gielesetal2011}. As also shown from Fig.~\ref{fig:fig1}, cluster orbital
parameters would not seem to be sufficient to hypothesize on the presence of tidal tails.

\begin{figure}
\includegraphics[width=\columnwidth]{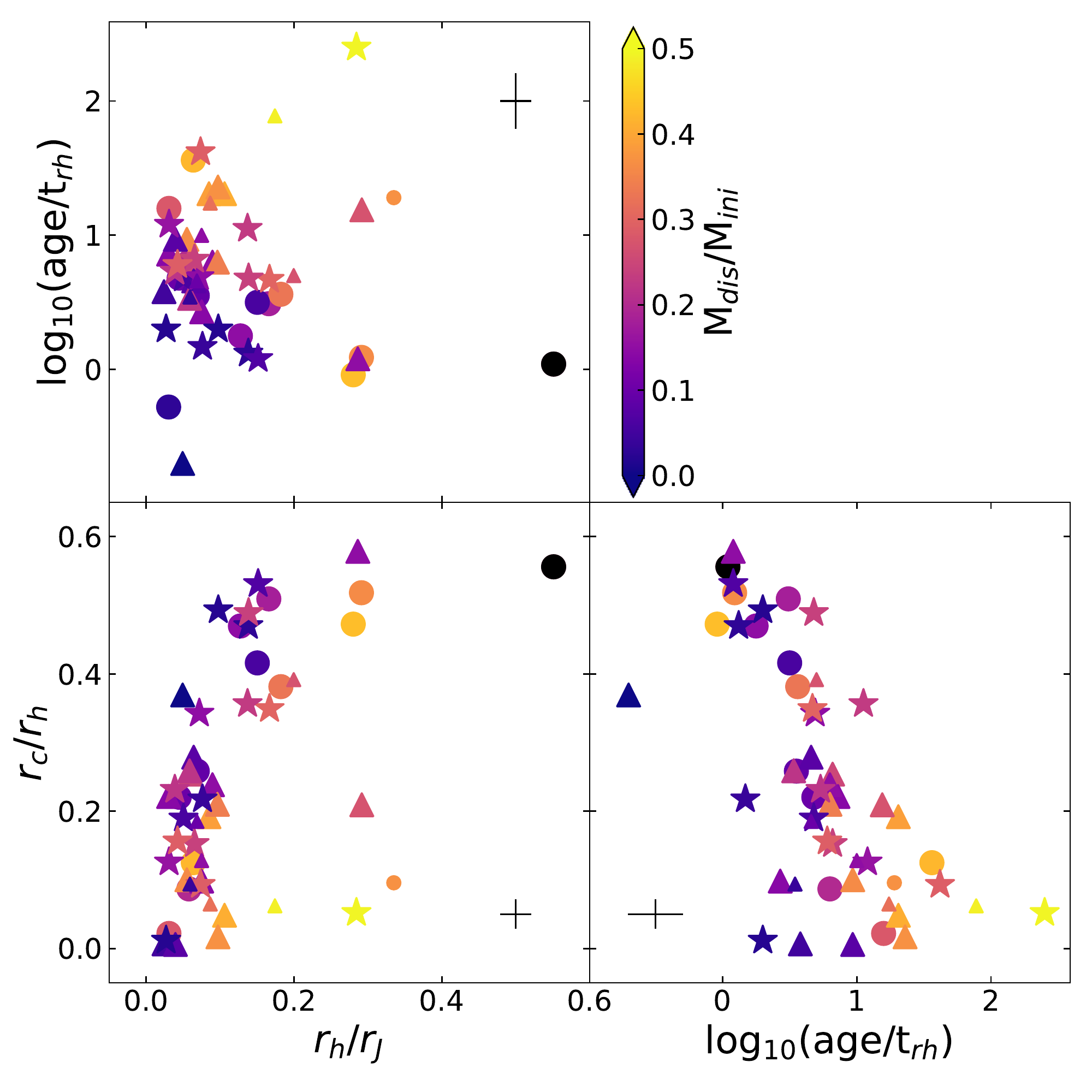}
\caption{Relationships between structural and internal dynamical properties
of globular clusters in groups G1, G2 and G3. Symbols are as in Fig.~\ref{fig:fig1}.
 Pal\,5 is represented by a large black filled circle.  Negative log(age/$t_h$)
come from adopting an average age for all globular clusters.}
\label{fig:fig3}
\end{figure}

We finally explored whether the presence of tidal tails makes any impact in shaping
structural and internal dynamical properties, such as core radius ($r_c$), half-mass 
radius ($r_h$) and the ratio of the age to the  half-mass relaxation time  \citep[time required 
for stars in a system to lose completely the memory of their initial velocity, $t_h$,][]{sh71}. 
This
is motivated by the fact that the mass-loss from which the tails are formed
could imply a change in the cluster stellar density profile and hence an advance
stage of its internal dynamical evolution \citep{pm2018,piattietal2019b}. 
We used the $r_c$, $r_h$ and $t_h$ values from  \citet{baumgardtetal2019}, where $t_{h}$ is calculated using the formalism of \citet{bh2018}. Globular cluster ages were assumed to be 12$^{-1.5}_{-2.0}$ Gyr  \citep{kruijssenetal2018}.
Fig.~\ref{fig:fig3} depicts several relationship, in which we represent globular
clusters in groups G1, G2 and G3 with the same symbols as in Fig.~\ref{fig:fig1}.
At first glance, globular clusters with observed tidal tails would not seem to
differentiate from those with \citet{king62} profiles. There is a general
trend that run for all three groups of clusters in Table~\ref{tab:tab1} in the
sense that: 1) clusters that have lost relatively more mass by disruption
do  not seem to have  preferentially inflated main bodies, 
although some highly disrupted clusters can be seen at 
$r_h/r_J \ga$ 0.2 (see left-hand
panels); 2) from them, those that have relatively more compact cores ($r_c/r_h
\la$  0.2) would seem to be in a more advanced stage of internal dynamical
evolution (log(age/$t_h$) $\ga$ 1.0,  where age/$t_h$ is a measure of how many
times the relaxation time a cluster has lived) (see bottom-right panel); and 3) those
globular clusters that have relatively expanded cores ($r_c/r_h \ga$ 0.4) or
relatively small main bodies ($r_h/r_J \la$ 0.2) would seem to be
in relatively less advanced stage of dynamical evolution (log(age/$t_h$) $\la$ 0.7) 
(see bottom panels).

\section{Summary and concluding remarks}

The presence of tidal tails in Milky Way globular clusters are witnesses of the interaction
experienced by them with their environment while traveling across the Galaxy. 
Since tidal tails are observed only in some globular clusters, we embarked in an 
observational-based analysis with the aim of tackling the conditions for whether or not a 
globular cluster can develop tidal tails.

For this purpose, we gathered from the literature reliable information about the existence
of tidal tails, and of other kind of extra-tidal structures, such as extended
low density halos, azimuthally irregular clumpy features, etc, and of satisfactory
fitting of \citet{king62} models to the outermost regions of the cluster stellar density profiles.
From this search we conclude, first of all, that not every globular cluster
in the Milky Way has tidal tails. Indeed, from 53 globular clusters included in our
final compilation, 14 have observed tidal tails,  22 have extra-tidal features different
from tidal tails and 17 present undetectable signatures
of extra-tidal structures. 

When exploring kinematical properties (orbit's eccentricity, inclination and semi-major axis)
in combination with the ratio of mass lost by disruption to the initial cluster mass, we 
found that there is no obvious clues to differentiate globular clusters with and without 
tidal tails.  All three defined groups of clusters (G1 for tails, G2 for extra-tidal features 
and G3 for \citet{king62} profiles) have similar kinematical properties.  In general, 
globular clusters moving in orbits with
a relatively high eccentricity (\ga 0.8) and with very inclined orbits ($\approx$ 
$\pm$70$\degr$ from the Milky Way plane) have lost relatively more mass due to tidal
disruption than those in more circular and less inclined orbits. 

We also found that globular clusters with larger apogalactic distances and smaller
remaining fraction of cluster mass than Pal\,5 -a very well known globular clusters with a
long tidal tail  highlighted by \citet{bg2018}-, are not necessarily candidates for developing tidal tails.
Furthermore, globular clusters with observed tidal tails are found to keep larger
fraction of remaining clusters mass and have smaller apogalactic distances than Pal\,5.
Additionally, globular clusters with extra-tidal features or \citet{king62} profiles
also span similar ranges of values in the $R_{apo}$ versus $\mu$ plane. We checked
that the initial mass is not correlated with the presence of tidal tails.

Finally, we investigated whether the internal dynamical evolution of globular clusters
are reached by the effect of escaping stars in the form of tidal tails. To this respect, we
considered different relationships between the core, half-mass and Jacobi radii, the
ratio of the cluster age to the respective relaxation time and the ratio of the mass
lost due to disruption to the total cluster mass. The outcomes show that irrespective of
the presence or the absence of any kind of extra-tidal characteristics, the globular
clusters can reach an advanced stage of their internal dynamical evolution if they
have lost a relatively large amount of mass by tidal disruption. Therefore, there
would seem that there is not any overall property that
allows us to predict the presence of tidal tails emerging for a given globular
cluster in the Milky Way.

\begin{acknowledgements}
We thank the referee for the thorough reading of the manuscript and
timely suggestions to improve it. 
A.E.P. acknowledge support from the Ministerio de Ciencia, Tecnolog\'{\i}a e Innovaci\'on Productiva (MINCyT) through grant PICT-201-0030. 
JAC-B acknowledges financial support to CAS-CONICYT 17003.

\end{acknowledgements}



\end{document}